\documentclass{osa-article}

\journal{oe}

\articletype{Research Article}

\begin{document}

\title{{Neural-adjoint method for the inverse design of all-dielectric metasurfaces}}

\author{Yang Deng,\authormark{1} Simiao Ren,\authormark{1} Kebin Fan,\authormark{1} Jordan M. Malof,\authormark{1} and Willie J. Padilla\authormark{1,*}}

\address{\authormark{1}Department of Electrical and Computer Engineering, Duke University, Durham, North Carolina 27708,
USA}

\email{\authormark{*}willie.padilla@duke.edu} 



\begin{abstract}
All-dielectric metasurfaces exhibit exotic electromagnetic responses, similar to those obtained with metal-based metamaterials. Research in all-dielectric metasurfaces currently uses relatively simple unit-cell designs, but increased geometrical complexity may yield even greater scattering states. Although machine learning has recently been applied to the design of metasurfaces with impressive results, the much more challenging task of finding a geometry that yields the desired spectra remains largely unsolved.  We explore and adapt a recent deep learning approach -- termed neural-adjoint -- and find it is capable of accurately and efficiently estimating complex geometry needed to yield a targeted frequency-dependent scattering. We also show how the neural-adjoint method can intelligently grow the design search space to include designs that increasingly and accurately approximate the desired scattering response. The neural-adjoint method is not restricted to the case demonstrated and may be applied to plasmonics, photonic bandgap, and other structured material systems.
\end{abstract}

\section{Introduction}
Electromagnetic metamaterials have been used to realize many exotic scattering responses over the last two decades. Effects including negative refractive index and cloaking have generated significant interest and have served to drive the community \cite{smith2000composite, pendry2000negative, shelby2001experimental, smith2004metamaterials, schurig2006metamaterial}. A more applied, but still relevant metamaterials achievement, is that of graded designs \cite{greegor2005simulation, smith2005gradient}. It was shown early on that unit-cells which form metamaterials may be designed with a spatial dependence across a surface and / or volume, and various lensing effects were shown that utilize spatial degrees of freedom. A principle of gradient metasurfaces is that their scattering properties change slowly as a function of spatial coordinate. Broadband metamaterial absorbers \cite{landy2008perfect, avitzour2009wide}, and metamaterial spatial light modulators \cite{shrekenhamer2013four}, also make use of dissimilar neighboring unit-cells, however there is no such requirement to make neighbors as alike as possible, and designs are simply cobbled together to achieve the desired response. This fact highlights a design feature of sub-wavelength metal-based metamaterials, i.e. their scattering response is primarily connected to their geometry and -- due to minimal neighbor interaction -- the unit-cell largely governs the electromagnetic properties of the array. Conventional optimization techniques built-in to modern day electromagnetic mode solvers are sufficient to achieve a desired response, and designs of differing units may be assembled with little change to the overall response.

In principle, all-dielectric metasurface (ADM) unit-cells may also be used to tesselate a surface in an arbitrary fashion similar to that achieved with metal-based metamaterials. However, ADMs resonators are only marginally sub-wavelength, and modes utilized are often not confined within their physical bound -- with an evanescent tail lying just outside their surface -- resulting in significant neighbor interaction. Further, surface modes related to the periodicity are not restricted to nearest neighbors, and may persist over several spatial periods. Despite the success of ADMs \cite{sautter2015active, headland2015terahertz}, it is conceivable that still-richer electromagnetic scattering can be achieved if more complex geometries are employed. However, physical understanding for such metasurfaces is poor -- simple functional relationships, or even heuristic guidance regarding super unit-cell geometry and final electromagnetic properties is unavailable. The only contemporary means to estimate such metasurface properties, given a candidate geometry, is simulation or fabrication.  However, given the vast space of potential designs and the speed of conventional simulation and fabrication, it is completely infeasible to iterate over designs in order to achieve a desired response. 

A viable design alternative to numerical simulation for structured materials, including metamaterials \cite{liu2018generative, ma2018deep, jiang2019global, nadell2019deep, campbell2019review}, photonic band gap \cite{peurifoy2018nanophotonic, liu2018training, kiarashinejad2020knowledge, huang2020deep}, and plasmonics studies \cite{inampudi2018neural, jiang2019free, wiecha2019deep}, is deep learning.  Deep neural networks (DNNs) have successfully learned a forward mapping $s=f(g)$ between a metasurface geometry $g$ and the resulting frequency dependent electromagnetic scattering $s$, where $f$ is an unknown complex (e.g., highly nonlinear) function. \cite{gareth2013introduction}  A DNN -- once trained -- can effectively act as a high-speed simulator that may be used to find the electromagnetic scattering of a candidate geometry substantially faster (e.g., by six orders of magnitude \cite{nadell2019deep}) than conventional simulation.

\section{Neural-Adjoint Method}
While deep learning enables substantially faster evaluation of candidates, given the vast number of possibilities for many problems (e.g., approx. $10^{12}$ in our case), it is nonetheless difficult or impossible to iterate over all or most candidate designs. From a design perspective, what would be of greatest utility would be to instead specify a desired frequency dependent electromagnetic response $s$, and have a model or solver compute a specific approximate geometry $\hat{g}$, which yields $s$ (here $\hat{}$ denotes approximate). The solution used to search for such an ideal geometry may be cast as an inverse problem written as $\hat{g}=\hat{f}^{-1}(s)$. Hadamard described a solution method as well-posed if it met three criteria -- $H_1$: existence (at least one solution), $H_2$: uniqueness (only one solution), and $H_3$: stability (solution must depend continuously on $s$). A solution approach is then ill-posed if one of the $H_1-H_3$ conditions fails \cite{mueller2012linear}. Past work has shown that finding a specific ADM consisting of distinct neighboring resonators in a super unit-cell geometry for a desired scattering state is an ill-posed non-linear inverse problem and, in particular, conditions $H_1$ and $H_2$ are not met \cite{liu2018generative}.

\begin{figure}[t!]
    \centering\includegraphics[width=0.5\textwidth]{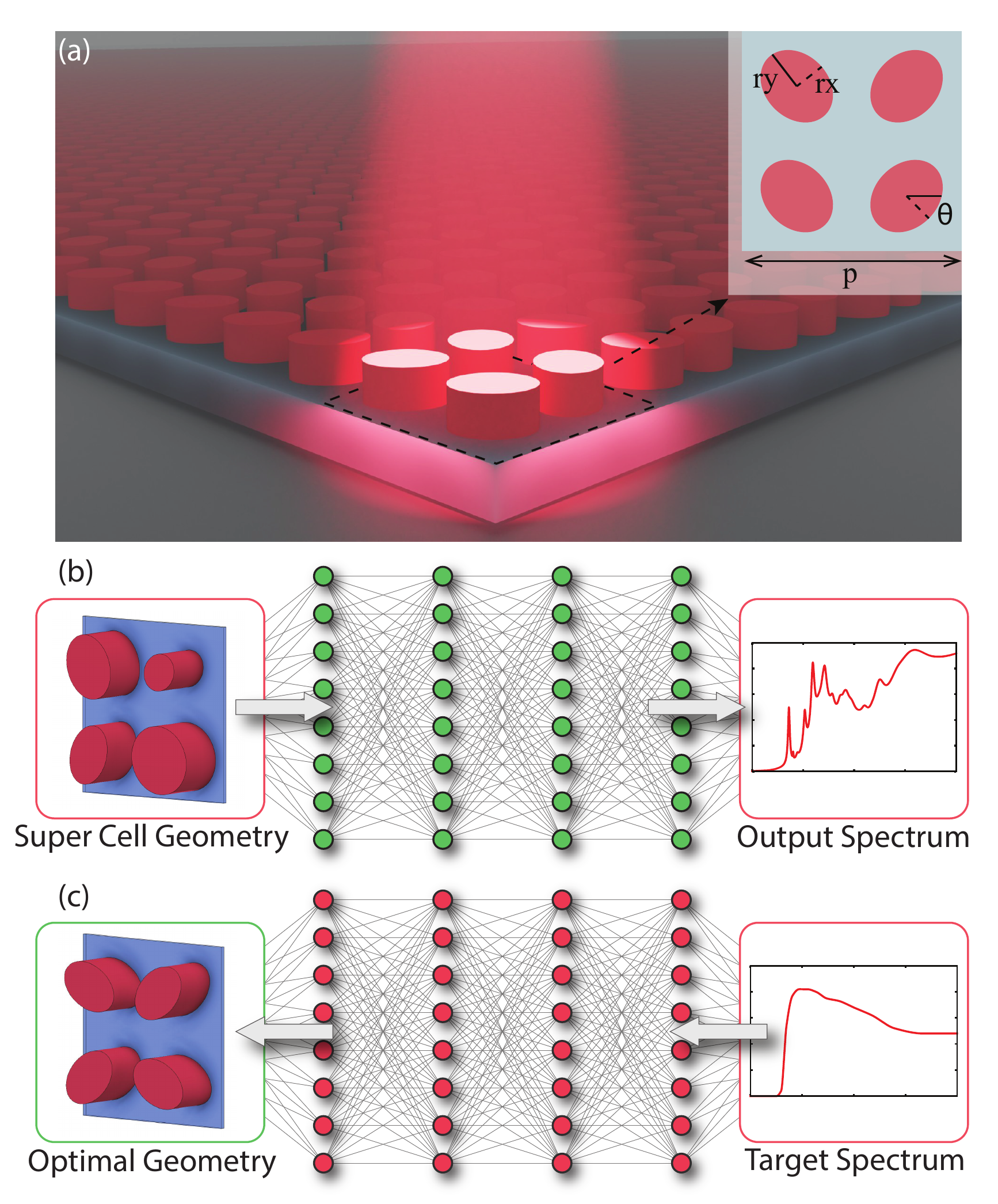}
    \caption{(a) 3D illustration of an all-dielectric metasurface absorber. A schematic of the all-dielectric metasurfaces super-cell is shown -- top right. (b) Training phase of the neural-adjoint method which uses fixed geometry inputs and corresponding simulated spectrum. (c) The inference phase of the neural-adjoint method, in which the target spectrum drives the search of the optimal geometry. The geometry is shown in a green box here to indicate it is the only trainable parameter.}
\label{1}
\end{figure}

Here we explore a recently-proposed deep inverse modeling approach, called the neural-adjoint (NA) method \cite{ren2020benchmarking}, which outperformed many other methods for solving ill-posed inverse problems.  Here we adapt the NA method to solve a challenging problem involving the design of ADMs geometries with 14 free geometrical parameters; a much larger space than most recent work (e.g., $5-10$ free parameters in  \cite{peurifoy2018nanophotonic, ma2018deep, nadell2019deep}).  Furthermore, existing inverse methods often involve complex training procedures, and ultimately produce sub-optimal solutions. In contrast, the NA only requires that we train a single conventional feed-forward neural network, and -- as we show -- appears to find close approximations to the globally optimal solution within just one minute, even for our complex ADM design problem.  Furthermore, we demonstrate how the NA method can be utilized to expand the design search space, essentially providing a form of active learning that is specifically tailored to solve inverse problems. We explore an example inverse problem, where a frequency dependent infrared absorptivity ($A(\omega)=s$) is desired, which we would like to achieve using an ADM consisting of a square array of 2$\times$2 resonators, each with an elliptical generatrix lying perpendicular to its directrix (line of length $h$) -- depicted in Fig. \ref{1}. Although past experience and single unit-cell simulations guide us to choose approximate metasurface dimensions which yield resonances in the desired spectral range, our inverse problem is ill-posed, since we do not know if a solution exists, i.e. condition $H_1$ is not met. Further, we do not meet the condition of uniqueness, $H_2$, since many metasurface solutions (disparate geometries) may yield the same spectra -- at least within the accuracy of our chosen loss metric, and numerical precision. The proposed ADMs consists of a geometry space of fourteen parameters: [$h$, $p$, $r_{x_1}$--$r_{x_4}$, $r_{y_1}$--$r_{y_4}$, $\theta_1$--$\theta_4$]. As shown in Fig. \ref{1}, $h$ is the height of all four elliptical resonators, $p$ is the periodicity of the super cell, $r_{x_1}$ -- $r_{x_4}$ and $r_{y_1}$ -- $r_{y_4}$ are the x-axis and y-axis radii of four elliptical resonators respectively, and $\theta_1$ -- $\theta_4$ are the rotational angles with respect to the center axis of each elliptical resonator. All geometry values are randomly sampled from the data grid shown in Table S1, which is included in the Supplemental. The numerical simulations use SiC for the proposed ADMs, and the Supplemental contains more details.

\begin{figure}[t!]
    \centering\includegraphics[width=\textwidth]{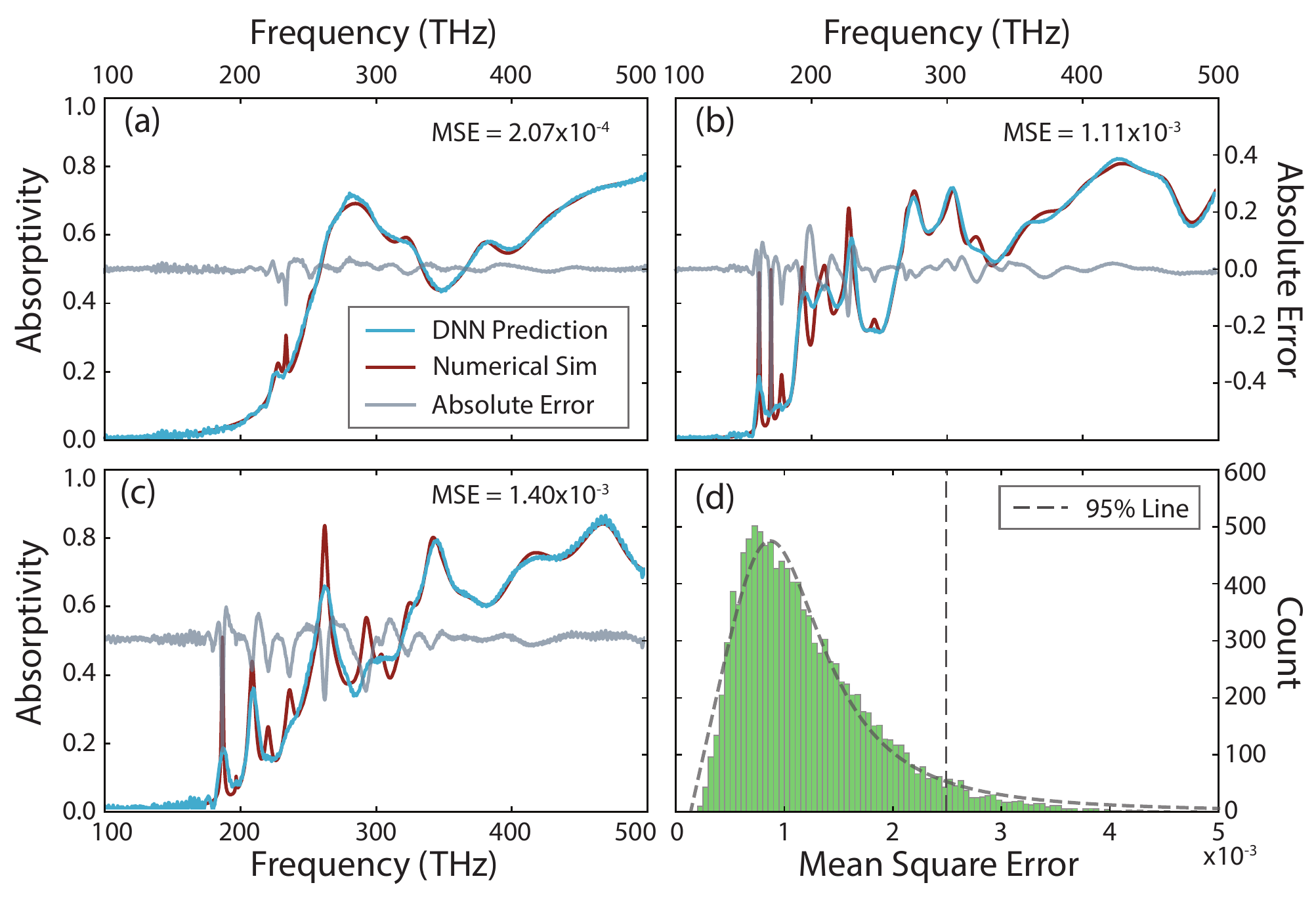}
    \caption{Forward model mean-squared-error (MSE) between DNN predicted spectra and numerical simulations. The DNN result with (a) the best, (b) average, and (c) below average performance. (d) Histogram detailing the distribution of MSE for the entire validation set.}
    \label{2}
\end{figure}

\section{Results}
The DNN forward model is trained by pair of 14 geometrical parameters and 2000 frequency points from 100-500 THz. Results of the forward model trained on 60k $\{g,s\}$ pairs (approximately $5.75\times10^{-8}$ of the entire geometry space) are shown in Fig. \ref{2} (a)-(c), where red curves are numerical simulations, and blue curves are the DNN predictions. The absolute error between simulation and prediction is plotted as the gray curve on the right vertical axis of Fig. \ref{2} (a)-(c). In Fig. \ref{2} (d) we show the histogram of the mean-squared-error (MSE) for the entire validation set, and find that 95\% have an MSE$\leq2.49\times10^{-3}$ (dashed gray line), and an average MSE of $1.2\times10^{-3}$. Having an accurate trained forward model we next turn toward the inverse solution. The NA method \cite{ren2020benchmarking} finds the optimal inverse solution by fixing all the weights and biases of $\hat{f}$, and computing the forward model's gradient solely with respect to the input to the network (i.e., the geometry), starting from randomly chosen values, denoted $\hat{g}_0$. It is important to highlight that $\hat{f}$ is a closed-form differentiable expression, and thus calculation of $\partial \hat{f}/\partial g$ is trivial. Further, we can estimate the gradient of the input geometry with respect to a loss function $\mathcal{L}$ that we are free to specify (e.g., mean squared error). Therefore, if $s$ is our desired spectrum, and $\hat{g}_i$ is our current best estimate of the metasurface geometry, we can iteratively move along the loss surface to find a better solution using, 

\begin{equation}
    \left. \hat{g}_{i+1} = \hat{g}_{i}+\alpha\frac{\partial \mathcal{L}(\hat{f}(\hat{g}_{i}),s)}{\partial g} \right\vert_{g = \hat{g}_{i}}\\
    \label{eq1}
\end{equation}

\noindent where $\alpha$ is the learning rate. The $\hat{g}_i$ can then be evaluated iteratively until some convergence criteria is satisfied (e.g., $\mathcal{L}$ changes very little after each iteration).  Because this is a gradient-based procedure, it will only converge to a locally optimal solution.  As a result, the NA method prescribes that this search process be repeated $T$ times, each time starting from a different randomly chosen value of $\hat{g}_{0}$. In practice we find that we may run greater than $10^4$ $T$ initializations in parallel with no cost in speed -- only limited by available memory. Therefore the NA method produces $T$ candidate designs, and we can choose the best design (or several designs, if desired) by passing each design back into $\hat{f}$ and evaluating their similarity to the target scattering properties.

\begin{figure}[t]
    \centering\includegraphics[width=\textwidth]{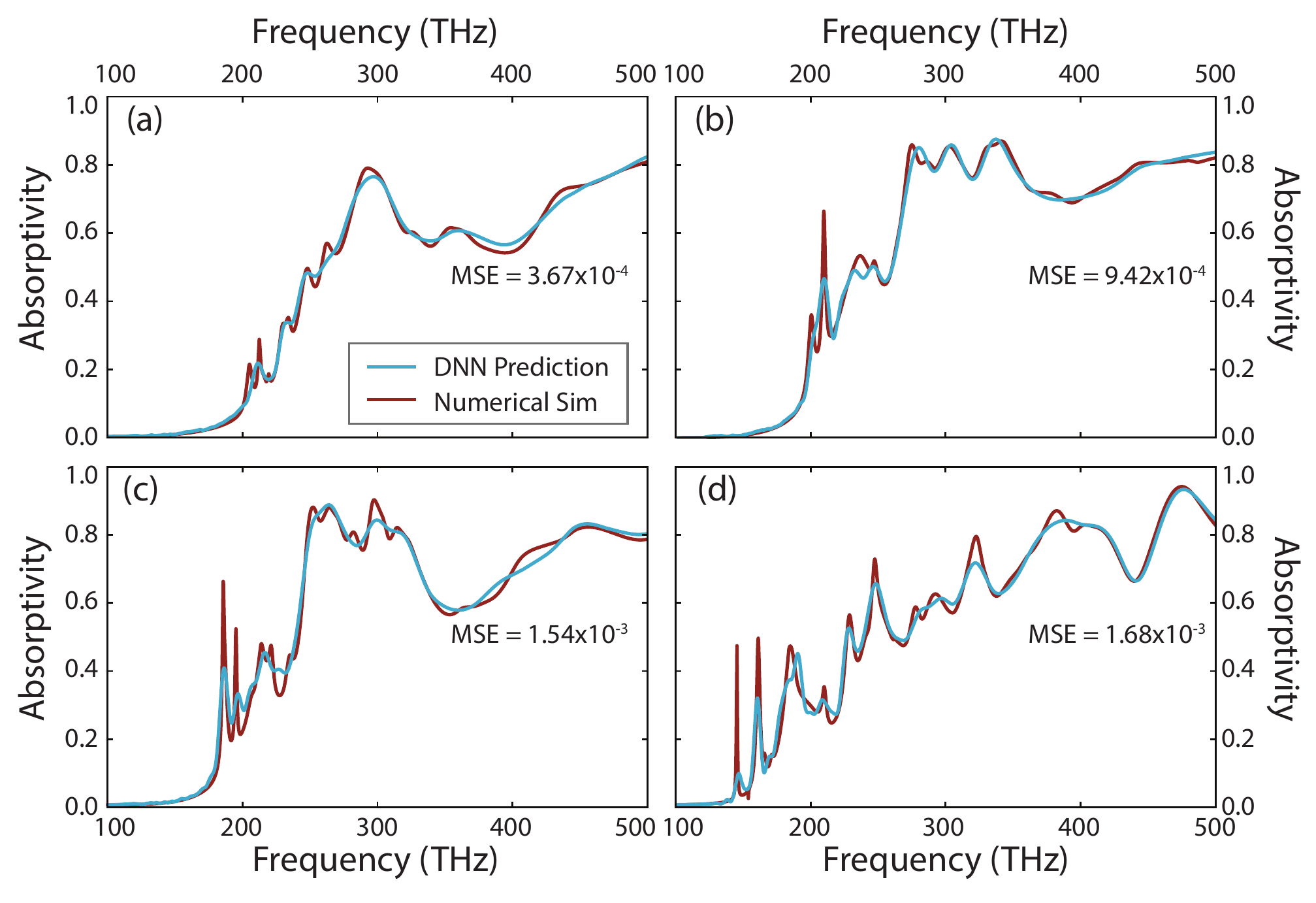}
    \caption{Neural adjoint inverse results for prediction of target spectra that exist within the geometry space. Example spectra from geometry predictions that have (a) the best, (b) average, and (c)-(d) below average performance compared to the target spectrum.}
\label{3}
\end{figure}

Notably, while the user can specify many different choices for $\mathcal{L}$, the NA method prescribes that a so-called "boundary loss", $\mathcal{L}_{bnd}$ should be added to any user-chosen loss \cite{ren2020benchmarking} and is given by: $\mathcal{L}_{bnd} = ReLU(|\hat{g}-\mu_g|-\frac{1}{2}R_g)$, where rectified linear unit (ReLU) is the activation function, $\mu_g$ is the mean of the geometry training data, and $R_g$ is its range.  This boundary loss punishes the inference process with increasing loss if the geometry search process steps out of the space of the training data, where the forward model may produce inaccurate estimates of scattering parameters.  In our experiments we use the following total loss: $\mathcal{L}= (s-\hat{f}(\hat{g}))^2 + \mathcal{L}_{bnd}$. As an initial test of the NA inverse method, we feed in frequency dependent absorptivities $A(\omega)$ where we know apriori that a solution $g$ exists, i.e. $s$ is a numerical simulation from which $\hat{f}$ originates from. In Fig. \ref{3} we show results of the NA inverse method and each sub-plot shows characteristic results of (a) the best results, (b) average results, and (c) and (d) below average results, all based on MSE.  In all of the these examples, the NA method identifies a close approximation to the correct solution.  This is impressive given the complexity of the spectra present in Fig. \ref{3}. We suspect the small remaining errors in the predicted design are due largely to the limited precision of gradient descent as it nears solutions (i.e., minima) in the error space; due to the non-zero learning rate in Eq. \ref{eq1}, it cannot converge to the exact minimum point.  We note however that learning rate can be gradually reduced during the search process, at the cost of additional computation time, until a solution of desired precision is obtained.

\begin{figure}[t]
    \centering\includegraphics[width=\textwidth]{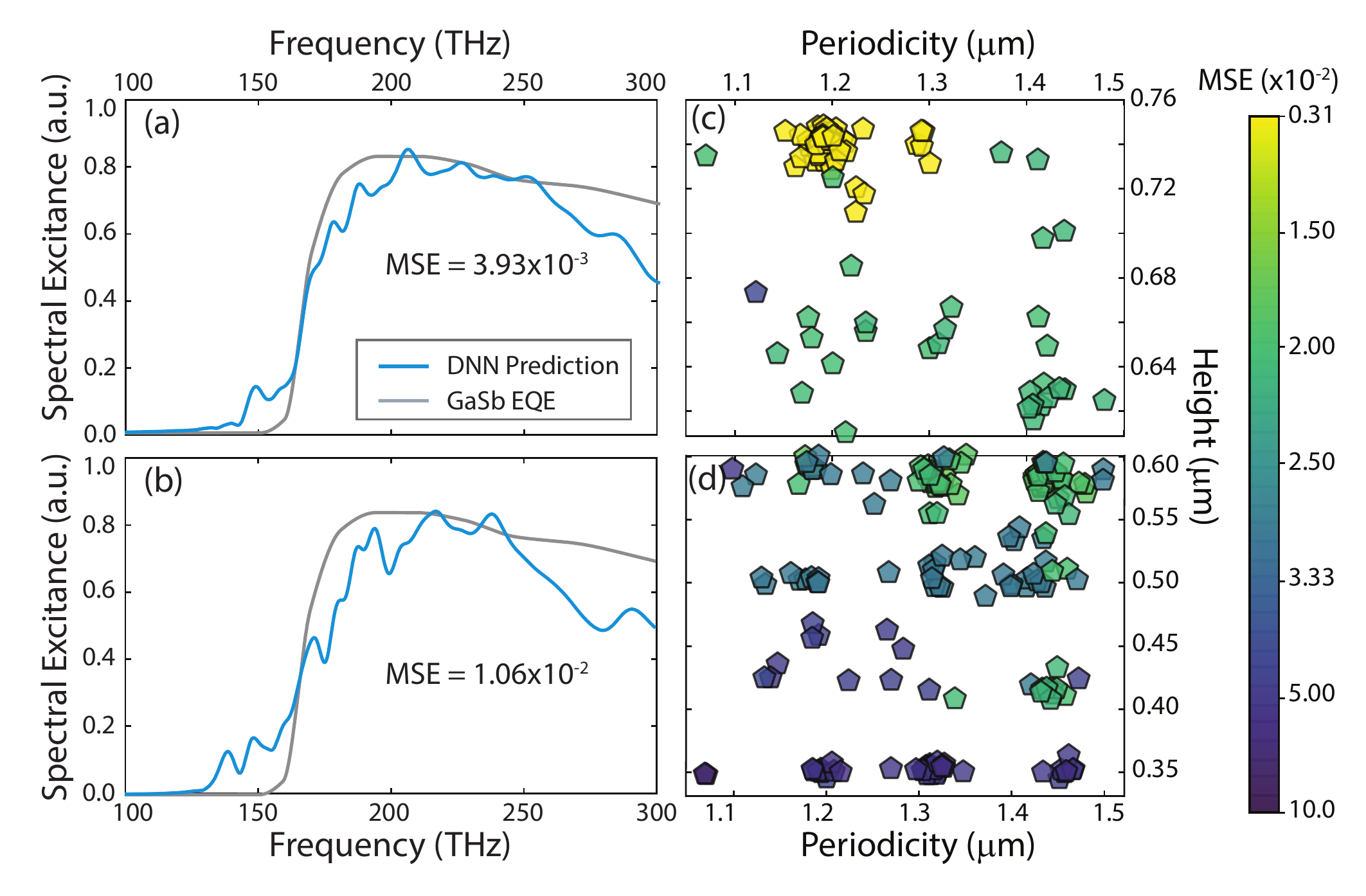}
    \caption{Neural adjoint inverse results for matching the EQE of GaSb. The target spectrum (gray), DNN prediction for $M_{e,\nu}(T=2100K)$ (blue), for the expanded geometry space (a), and  the original geometry space (b) explored. Neural adjoint predictions versus $p$ and $h$ for (c) the expanded geometry space, and (d) the original space. Symbol colors indicate MSE, with values given in the colorbar.}
\label{4}
\end{figure}

\begin{figure}[t]
   \centering\includegraphics[width=\textwidth]{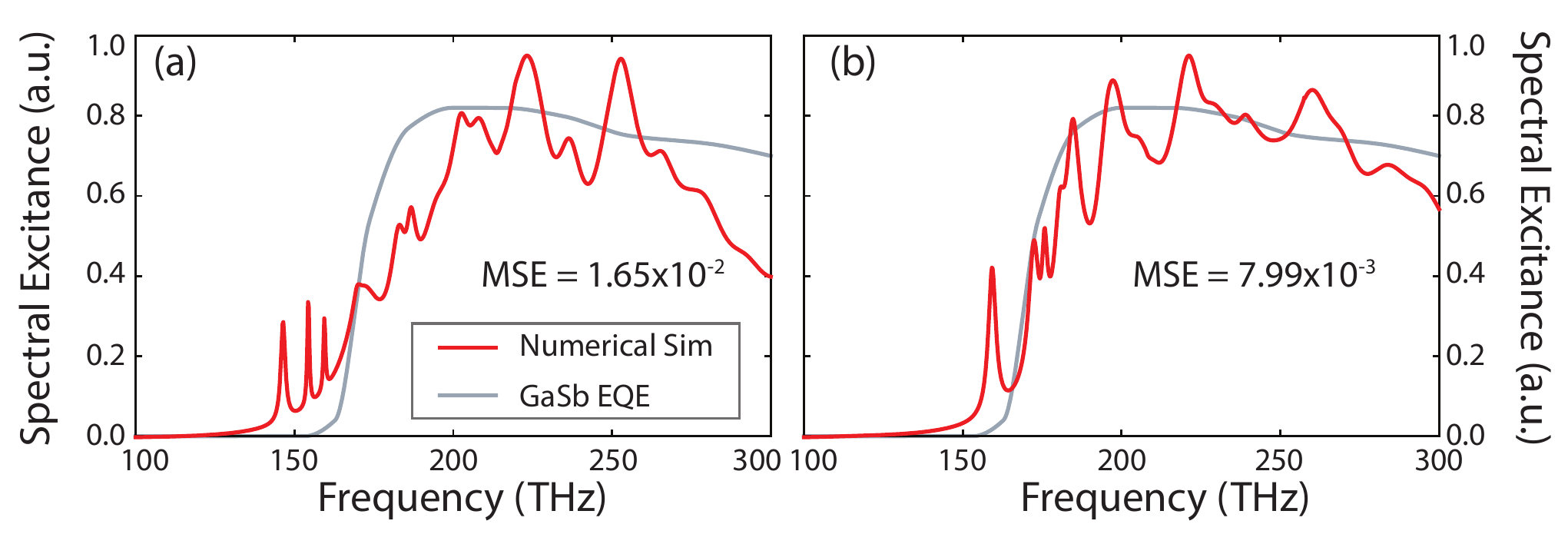}
    \caption{Numerically simulated $M_{e,\nu}(T=2100K)$ (red) of the optimal geometry predicted by the neural-adjoint method to match the EQE of GaSb (gray), for the original geometry space (a), and the expanded geometry space (b).}
\label{5}
\end{figure}

We next turn to the significantly more challenging task of applying the NA method to a spectra where we are unaware if a solution exists within the chosen geometrical parameter space, i.e. criterion $H_1$ for inverse problems is violated. We chose the frequency dependent external quantum efficiency (EQE) of gallium antimonide (GaSb) as $s$, shown as the gray curves of Fig. \ref{4}. The metasurface will operate at elevated temperature, and thus we consider the so-called graybody spectra --  also termed the spectral exitance $M_{e,\nu}(T)$ -- which is given by the blackbody radiation curve times the absorptivity. We keep the top 16,000 neural adjoint solutions (spectra) and determine $M_{e,\nu}(T)$ for each of these at 100 temperatures between 1500 and 2500 k --  a total of $1.6\times 10^6$ candidates. The shape of the EQE curve differs significantly from typical spectra we see in our geometry space (Fig. \ref{3}). None-the-less we find a best solution resulting from the NA method at a temperature of $T=2100k$ that consists of a geometry of [$h=0.566$, $p=1.440$, $r_{x_1}=0.180$, $r_{x_2}=0.155$, $r_{x_3}=0.214$, $r_{x_4}=0.278$, $r_{y_1}=0.285$, $r_{y_2}=0.253$, $r_{y_3}=0.146$, $r_{y_4}=0.256$, $\theta_1=-0.901^{\circ}$, $\theta_2=-20.677^{\circ}$, $\theta_3=-37.982^{\circ}$, and $\theta_4=39.046^{\circ}$]. The spectral exitance resulting from this geometry -- calculated from $\hat{f}$ -- is shown as the blue curve of Fig. \ref{4} (b), and we find an MSE, compared to the EQE of GaSb, of 1.06$\times$10$^{-2}$. We also apply a weighting function $W(\nu)=1\chi_{[100,275]}+0\chi_{(275,300]}$ on the MSE forcing the NA method to focus on the region of interest for energy harvesting purposes. To verify our neural adjoint results, we numerically simulate the predicted geometry and plot the resulting $M_{e,\nu}(T)$ in Fig. \ref{5} (a) as the red curve -- again compared to the EQE of GaSb (gray curve). As can be seen, the simulated curve has many relatively sharp peaks that are not present in Fig. \ref{4}(b). This is because, as noted earlier, $\hat{f}$ does not perfectly match the numerical simulator, and this will introduce errors in the design process. Thus since the NA method relies on $\hat{f}$ to search for designs, it is also limited by the accuracy of the forward model estimate. We also found that since $\hat{f}$ is trained from geometries constrained to a grid, the discrepancy between NA solutions and numerical simulation arises because NA solutions are not confined to the grid, where our model is most accurate. None-the-less we find that our simulated $M_{e,\nu}(T)$ achieves an MSE of  1.65$\times10^{-2}$, as shown in Fig. \ref{5} (a).

Another major obstacle is that our design search space does not contain a geometry that can realize our target spectrum (i.e., Hadamard's criteria $H_1$).  This is suggested by the fact that our best NA solution, shown in Fig. \ref{4}(a), still does not match our target spectrum.  However, we can use the NA output to identify where we should expand our search space so that it will include better designs. We can do this by visualizing the error of all inverse solutions returned by NA, and looking for trends e.g., we may find that all the best solutions are bunched up against some edge of our initial search space, suggesting that expanding along that dimension may yield better results. However, since we have a 14 dimensional design space, we are unable to easily visualize these data. To address this problem, we use the Uniform Manifold Approximation and Projection (UMAP) \cite{mcinnes2018umap}, which is a type of dimensionality reduction method permitting us to visualize the distribution of our inverse solutions performance in 2D, so that we may more easily identify patterns. From this initial investigation with UMAP in Fig. S1 -- shown in Supplemental -- we find that our best NA inverse solutions are limited by height. Shown in Fig. \ref{4} (d) are NA solutions as a function of height and periodicity color mapped by corresponding MSE values. It is evident that not only are our best solutions grouped at the maximum height allowed in our geometrical space, $h=0.6$ $\mu$m, but also that the solutions improve as a function of height. Encouraged by these results, we expanded our original geometry space to include increased height values from 0.6 to 0.75 $\mu$m, by simulating an additional 24k $\{g,s\}$ pairs. The NA model now trained on the expanded geometrical space indeed finds an improved solution, shown as the blue curve in Fig. \ref{4} (a), where we realize an MSE that is reduced by a factor of 2.7. The simulated red curve in Fig. \ref{5} (b) further validates the result that the MSE of numerical simulations  is also reduced -- here by a factor of 4.8. A plot of the 1000 best NA solutions in the expanded geometrical space shown in Fig. \ref{4} (c), however, indicate that we may be able to make continued improvements, since we still have a gradient pushing for greater heights -- although the periodicity seems to be honing in on a value of 1.2 $\mu$m.

\section{Conclusion}
We have adapted the neural-adjoint inverse design method \cite{ren2020benchmarking} to accurately predict the high-dimensional all-dielectric metasurface geometry needed to produce a targeted infrared absorptivity spectra. When the geometry needed to produce a desired spectrum lies outside of the bounds, the NA method appears to find the best possible solution within the permissible search space. Unlike other adjoint inverse approaches \cite{lalau2013adjoint}, the NA method does not require any domain knowledge specific to the problem. In the event that the inverse solution does not exist in the parameter space explored, NA may be used to guide one to a better solution, through expansion of the design parameter search space. This may help to reduce the initial required number of numerical simulations, and to instead use NA guided simulation exploration. With its exceptional computational speed, high accuracy, and potential use in active learning that is explored here, the neural-adjoint method has an impressive future in not only ADMs thermal emitter but also any ADMs inverse problems. The NA method is not restricted to the case presented, but may be applied to many other systems including photonic band gap and plasmonics.


\bibliography{ML_MM_reference.bib}

\end{document}


\maketitle

\section{Numerical Simulation}
The cylindrical resonators' geometry was previously demonstrated in the THz regime by\cite{liu2017experimental, fan2017all}. To prove deep neural work capability with high-dimensional inputs, we increased the geometrical dimensionality by introducing the elliptical structures to previous cylindrical resonators, and each elliptical resonator undergoes a rotation angle ranging from -45 to 45 degrees. Furthermore, governing the fabrication practicality, we fixed all elliptical resonators to have the same height. To migrate from the THz regime to the infrared, materials and geometry sizes are scaled accordingly. We sized down our unit cell volume refers to the ratio of THz frequency and infrared frequency used in the legacy design and current design. Then an optimization on an adequate scale to finalize our geometry boundary listed in Table 1. We chose SiC for our simulations, considering its high melting point, high oxidation resistance, and reasonable absorption coefficient in near-infrared. We used experimentally measured relative permittivity data of SiC from 0 to 300 THz to fit the dispersive materials property of SiC in our simulations. To implement the unit cell boundary conditions in CST Microwave Studio we used for our simulation, we used finite frequency solvers to perform the numerical simulations, which also take considerations of the coupling effect between the four resonators within the unit cell. To minimize the time cost of the simulations, we lightly comprise the simulation accuracy to simulation speed. We tuned the Floquent mode to have one mode at both ports. The simulation mesh is tetrahedral, and we used a second-order solver with accuracy at $1e^-6$. The resulting spectrum of the simulation has 2001 data points within the range of 100-500 THz.

\begin{table}[htbp!]
\caption{Grid definition for the 14-dimensional input geometry parameters. h, p, and r are in units of microns. $\theta$ is in unit of degrees.}
\centering\begin{tabular}{ccccc}
 \\
 Step&h&p&$r_{x_n}$/$r_{y_n}$&$\theta_n$\tnote{*}\\ \hline
 1&0.3&1&0.1&-45 \\
 2&0.375&1.125&0.1125&-22.5 \\
 3&0.45&1.25&0.125&0 \\
 4&0.525&1.375&0.1375&22.5 \\
 5&0.6&1.5&0.15&45 \\
 6&-&-&0.1625&- \\
 7&-&-&0.175&- \\
 8&-&-&0.1875&- \\
 9&-&-&0.25&- \\
 \hline
\end{tabular}
\begin{tablenotes}\footnotesize
\centering\item[*] $n$ corresponds to the first to the fourth elliptical resonator in one super cell
\end{tablenotes}
\end{table}

The total possible number of geometrical combinations of our 2$\times$2 metasurface is $8^9*6^5=1.04\times 10^{12}$. It is impossible to use a conventional numerical simulation approach to exploit the entire geometry space to achieve the targeted spectrum. We find that the average simulation time per geometrical configuration per CPU is approximately three minutes. Thus it would take about 600 million years to finish exploring the entire geometry space with one CPU. The fast forward dictionary search (FFDS) inverse method was shown feasible for THz ADMs absorbers, where all 812 million possible geometries can be computed in a day \cite{nadell2019deep}. To compute our entire geometry space with a size of over a trillion parameters would take FFDS over three and a half years. Thus the NA method is a good choice when the parameter space become too large for a FFDS approach. 

\section{Deep Neural Network Architecture}
We built the entire network and neural-adjoint method using the PyTorch platform. The DNN used for the neural adjoint method consists of twelve fully connected linear layers, four 1D transpose convolutional layers for upsampling, and one final 1D convolutional layer for spectrum smoothing. The linear fully connected layers have the following structure[14, 1000, 1000, 1000, 1000, 1000, 1000, 1000, 1000, 1000, 1000, 1000, 1000, 500
], and all hidden layers except the last linear layer are batch normalized, activated by Leaky\_Relu, followed by dropout layers with p = 0.05. The transpose convolutional layers have kernel size [16, 16, 33, 33] and filter size [4, 4, 4, 4]. The final convolutional layer has kernel size 1 and stride at 1. The data loader takes geometry inputs and the first 2000 data points of the absorptivity spectrum to generate the training and test datasets.  The post-processing truncates the predicted spectra's first and last fifty points to drop the convolutional layers bump at the edges.  We use L2 regularization, batch normalization, and the ADAM optimizer.

\section{NA Inverse Method}
Because NA a gradient-based procedure, it will only converge to a locally optimal solution. Because the NA method is amenable to parallelization on graphics processing units (standard hardware for deep learning), this entire process can be computed in very little time.  In our experiments, we can run the NA method with $T=1000$ on our desktop computer with an Nvidia 2080ti GPU and complete processing in under 1 minute.Since the NA method finds the globally optimal solution even in its worst-performing cases, our results suggest that the NA always (or nearly always) finds the globally optimal solutions, even for highly complex problems like ours.  Interestingly, this suggests that the main obstacle of custom design is no longer the inverse model, but rather the space over which we choose to search for designs i.e., the shapes we consider (cylinder, crosses, etc) and their parameter settings (e.g., radii, height).  Although powerful, to use deep learning methods we must necessarily define a range of these settings so that we can collect simulations to train our models, and this space limits where we can search for designs.  However, the design needed to realize our targeted scattering parameters may not exist in this initial search space.   As we show subsequently, the NA method can also be used to identify where this initial search space can be expanded so that it is most likely to contain the desired solution, providing a solution to this emergent obstacle to complex material design.    

\section{Data Augmentation}
Our simulations' unit cell boundary conditions allow us to do four times data augmentation on our dataset because the infinite plane of unit cells consists of four different resonators' combinations that give almost identical spectra with fluctuations from CST software. However, we learned that the DNN could quickly learn the correlation between four different resonators' combinations. The forward model will know which input geometries share the same spectrum in high fidelity if the entire dataset is augmented before splitting into the training and validation dataset. Therefore, the forward model will give a false mean square error much lower than actual loss performance on an independent validation set. After applying data augmentation to the training and validation sets after the splits, we observe that the four times augmentation did not significantly improve accuracy. We believe that, with our 60000 simulations (24000 simulations after augmentation), the augmented data points are still too sparse to cover the entire geometry space defined by our geometry boundaries. 

\section{Geometry Space Exploration Through UMAP}
We use Uniform Manifold Approximation and Projection (UMAP) to explore our solution geometry space and realize that angles have more random impacts on distributing the best NA solutions.  Thus, we plotted the UMAP with ten parameters, excluding rotational angles. The plotted UMAP demonstrates a clear trend that the MSE is decreasing in one direction. To confirm that the decreasing trend matches the increasing of resonators' height, we further marked the points from maximum and minimum height boundaries, respectively. The clustering of points towards the best MSE performance suggests that the NA method is finding the best local minima. 

\begin{figure}[h!]
    \centering\includegraphics[width=3.0in]{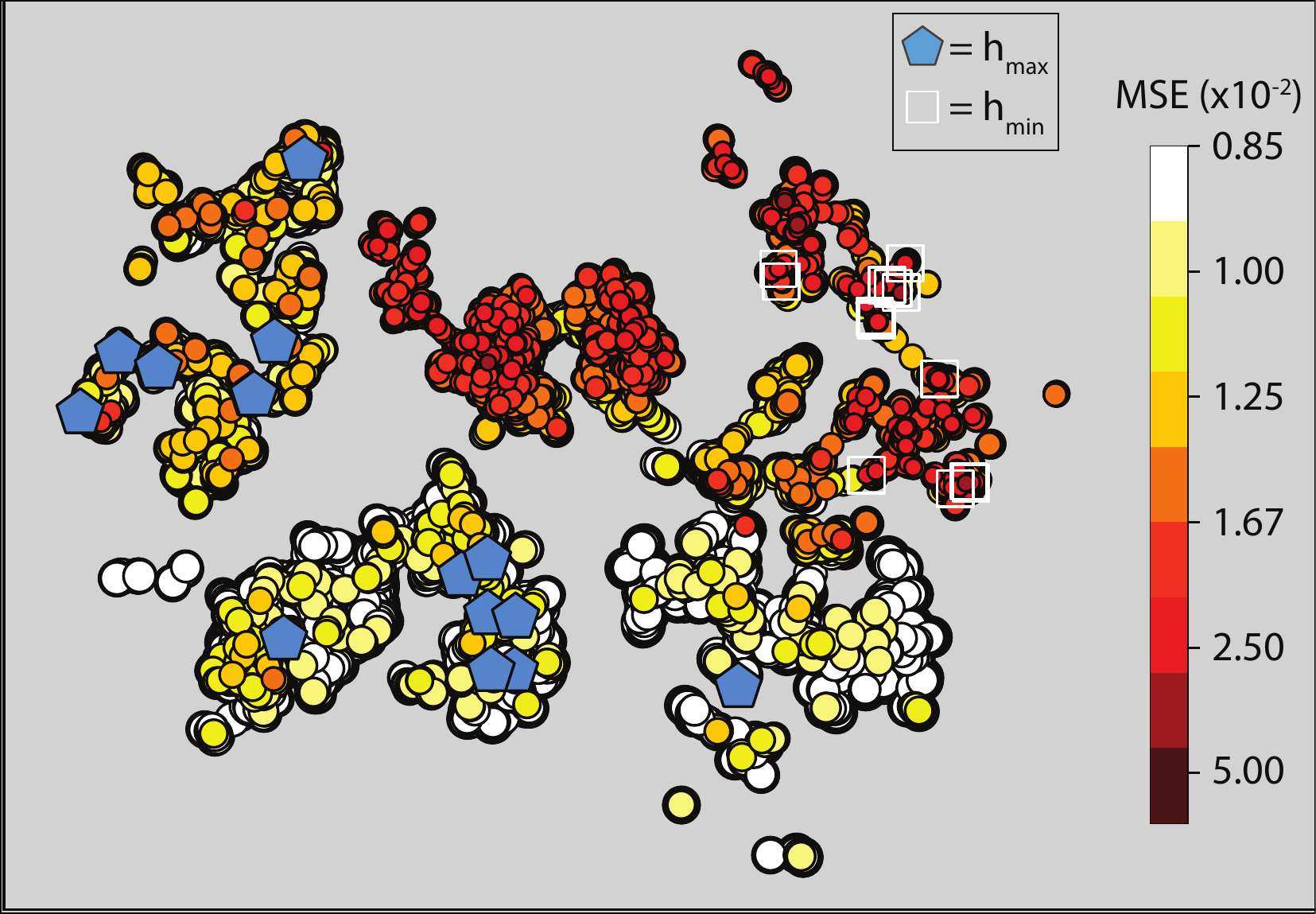}
    \caption{Uniform Manifold Approximation and Projection plotted with 10-dimensional geometry inputs indicates a strong correlation between the MSE performance and the increasing height.}
\label{5}
\end{figure}

\bibliography{ML_MM_reference.bib} 